\documentclass[11pt]{article}

\usepackage[a4paper,top=2cm,bottom=2cm,left=2cm,right=2cm,marginparwidth=0cm]{geometry}

\usepackage{amsmath}
\usepackage{graphicx}
\usepackage{wrapfig}
\usepackage{titling}
\usepackage{fancyhdr}
\usepackage{lastpage}
\usepackage{etoolbox}
\usepackage{ulem}
\usepackage{setspace}

\usepackage[table]{xcolor}

\definecolor{lightyellow}{rgb}{1.0, 1.0, 0.8}
\definecolor{softyellow}{rgb}{1.0, 0.9, 0.5}
\definecolor{deepyellow}{rgb}{1.0, 0.8, 0.0}
\definecolor{gold}{rgb}{1.0, 0.84, 0.0}

\usepackage{wrapfig}

\usepackage[colorlinks=true, allcolors=blue]{hyperref}

\usepackage[T1]{fontenc}
\usepackage{times}

\title{Exploring shell effects in fission yields of neutron-deficient\\ Th, Ac, and Ra isotopes near N=126}
\date{}

\begin{document}
\maketitle

\vspace{-3cm}
\begin{center}
{
J.~L.~Rodr\'{i}guez-S\'{a}nchez$^{1,2}$(Spokesperson), J.~Taïeb$^{3,4}$(Co-Spokesperson) 
for the R$^{3}$B collaboration
}
\end{center}

\noindent
\begin{small}
\textit{
$^{1}$CITENI, University of Coru\~{n}a, Ferrol, Spain 
$^{2}$IGFAE, University of Santiago de Compostela, Santiago de Compostela, Spain 
$^{3}$CEA/DAM/DIF, Arpajon, France 
$^{4}$Université Paris-Saclay, CEA, Bruyères-le-Châtel, France 
}
\end{small}

\begin{abstract}
Studies of nuclear fission over recent decades have led to a well-defined mapping of neutron and proton shell effects across the nuclear chart, particularly within the valley of stability. These shell effects play a crucial role in driving the asymmetric splitting of fissioning nuclei, as reflected in fission yields that are strongly influenced by spherical and deformed shell closures. In the actinide region, the existence of two primary fission modes, standard I and standard II, has been well-established. These fission modes are associated with the proton (neutron) shells at $Z=52$ ($N=82$) and $Z=56$ ($N=88$), respectively. Recently, a new proton shell around $Z=36$ has been found for the lighter fission fragments of pre-actinide nuclei. This discovery demonstrates that as we expand fission studies towards more exotic regions of the nuclear chart, new shell structures emerge. In this proposal, we aim to explore for the first time the most neutron-deficient isotopes of Th, Ac, and Ra. This region offers a unique opportunity to investigate stabilization effects around the spherical neutron shell $N=50$. To achieve this, we plan to use a primary beam of $^{238}$U at 1~GeV/u together with the Fragment Separator (FRS) to produce secondary beams of $^{213-216}$Th, $^{209-214}$Ac and $^{207-213}$Ra. For the investigation of the fission process, we will use the experimental methodology successfully applied in the S415, S438, and S455 experiments, being a continuation of those studies. Fission will be induced by electromagnetic-excitation (Coulex) reactions in inverse kinematics on an active target composed of Pb and C foils. The resulting fission fragments, together with emitted neutrons, will be measured using the R$^3$B experimental setup, which allows for complete kinematic measurements. The fission fragments will be identified in terms of charge ($\Delta Z$~<~0.34~FWHM) and mass ($\Delta A$~<~0.8~FWHM) while also determining their total kinetic energies~(TKEs). Additionally, neutron multiplicities will be measured using the NeuLAND detector.
\end{abstract}

\section{Introduction}
Nuclear fission, despite having been discovered over 87 years ago, continues to be a complex and enigmatic process. This phenomenon, involving the splitting of a heavy nucleus into two fragments after undergoing extreme deformations, is accompanied by the release of substantial amounts of energy and particle emission. The study of nuclear fission is pivotal across diverse fields, including nuclear physics, astrophysics, applied sciences, and other domains where quantum mechanics and non-equilibrium dynamics constitute the theoretical foundation. The inherent complexities of fission arise from the interplay between intrinsic and collective degrees of freedom, governed by the coupling of individual particle motion with large-scale nuclear deformations. These dynamics are strongly affected by quantum shell effects and correlations, particularly at low excitation energies~\cite{Schmidt2018}, which makes the phenomenon a valuable laboratory for advancing our understanding of nuclear structure and collective dynamics.

\begin{figure}[h!]
    \centering
    \includegraphics[width=0.565\linewidth]{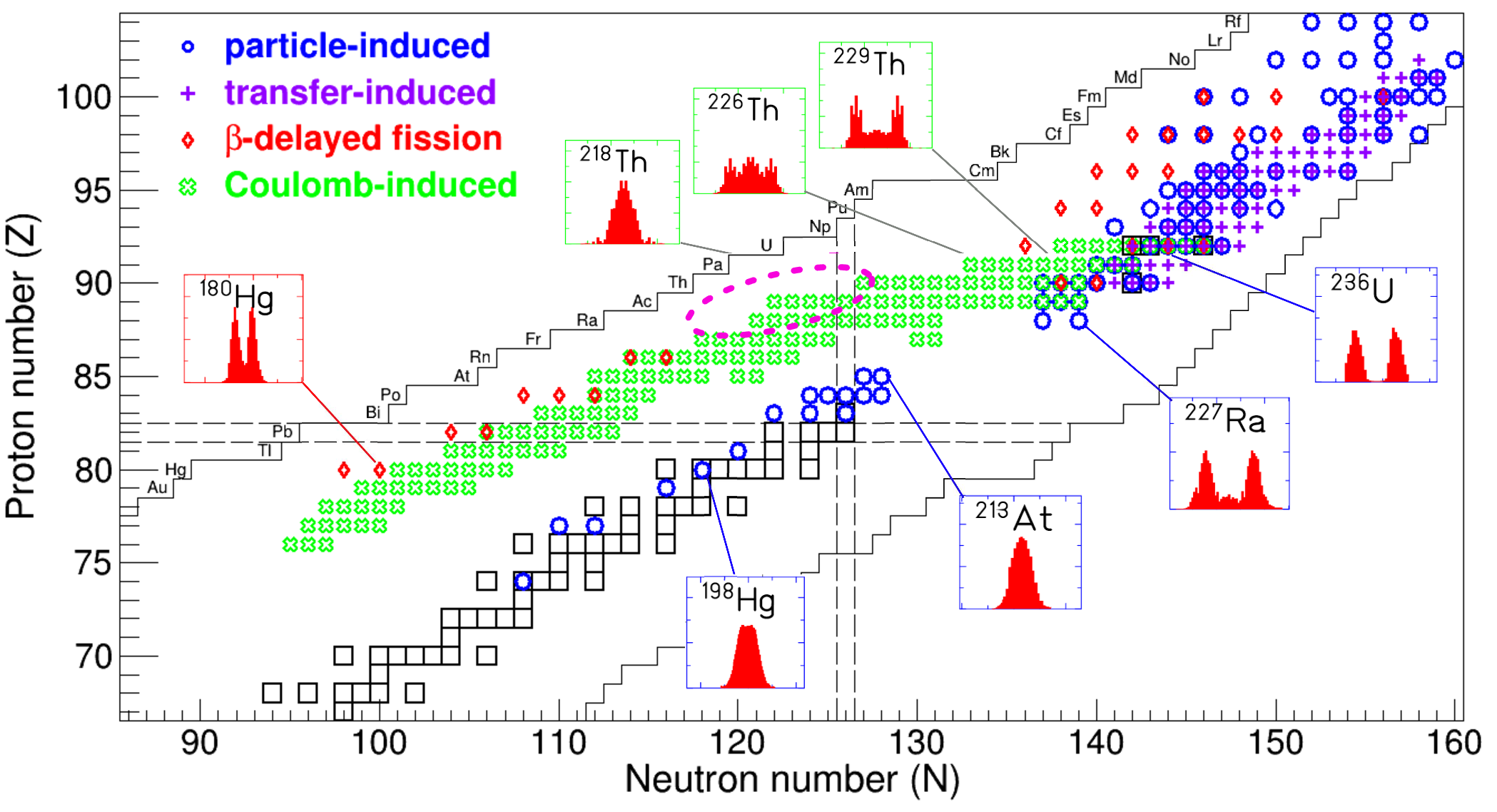}
    \includegraphics[width=0.425\linewidth]{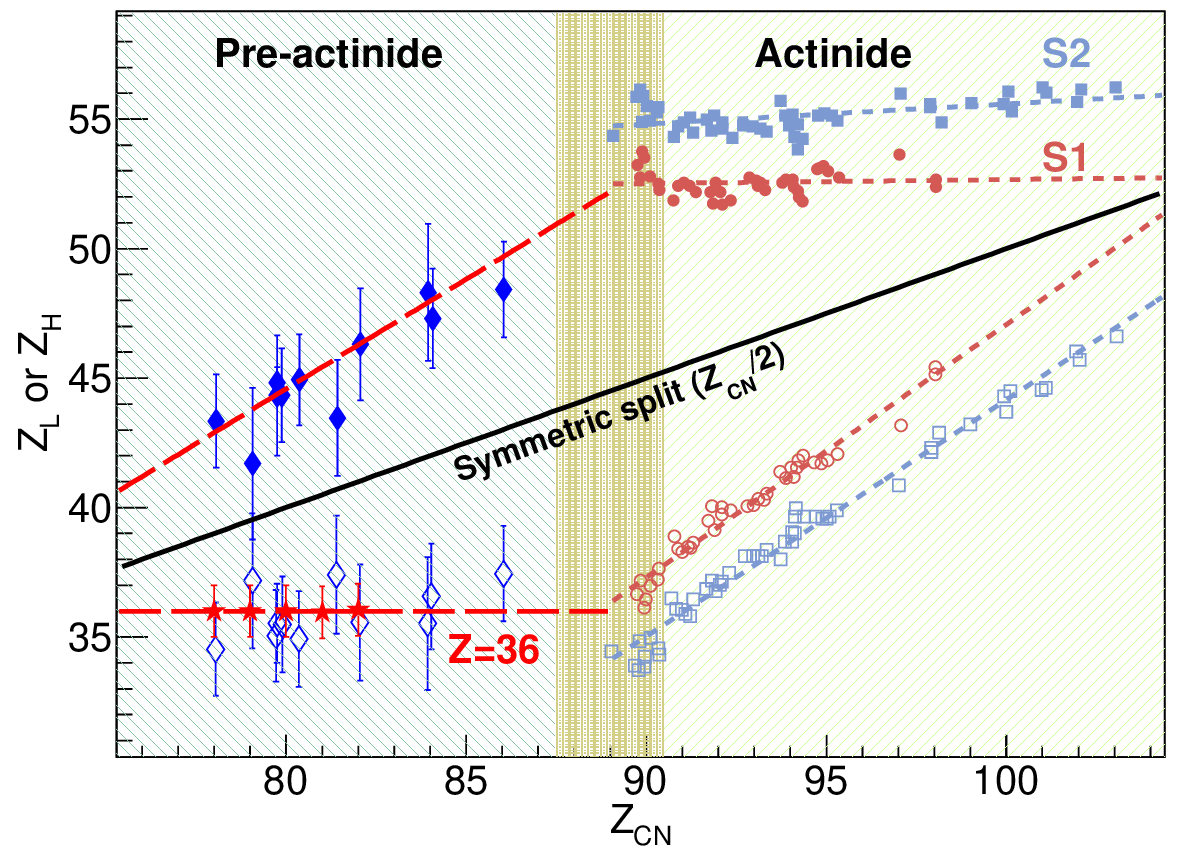}
    \caption{(Left) Overview of fissioning systems investigated up to 2024 in low-energy fission with excitation energies up to $\sim$15~MeV above the fission barrier~\cite{Aumann2024}. The pink ellipse indicates the region that will be investigated with this experiment. (Right) Evolution of the position of the standard I (S1, $Z_H$=52) and II (S2, $Z_H$=56) fission modes~\cite{Mahata2022} with the atomic number of the fissioning system (Z$_{CN}$), as well as the new stabilization around $Z_L \sim$ 36 observed in the pre-actinide region. The shaded region spanning $Z_{CN}$ from 88 to 90 marks the area of investigation for this proposal.
    }
    \label{fig:1}
\end{figure}

The shell effects, spherical and deformed, manifest in the charge and mass fission yields as well as in the TKE distributions. The interplay among different proton and neutron shells arises distinct fission modes, such as the well-defined super-long symmetric channel and the asymmetric modes Standard I (S1) and Standard II (S2) associated with the proton (neutron) shells $Z=52$ ($N=82$) and $Z=56$ ($N=88$), respectively. An overview of the fission-fragment distributions measured in the last two decades at GSI is shown in the Fig.~\ref{fig:1} (left) together with other measurements~\cite{Aumann2024} based on particle-induced and spontaneous fission (blue circles), $\beta$-delayed fission (red diamonds), and transfer-induced fission reactions (plus symbols). The most significant result is the transition from asymmetric to symmetric fission observed in the Th-Ra region, as well as the reverse transition from symmetric to asymmetric fission for fissioning nuclei below At. These phenomena result in two distinct islands of asymmetry, the origins of which remain not yet well understood. On the one hand, Ichikawa and M{\"o}ller~\cite{Ichikawa19,Moller2015}, using a macroscopic-microscopic approach, attribute the observed mass asymmetry below At to a shell gap developing at the outer saddle point in the neutron sub-systems. 
On the other hand, Scamps and Simenel, employing a microscopic energy density functional framework~\cite{Scamps19}, have attributed this asymmetry to the dominance of octupole effects, which are often driven by neutron configurations.

The pioneering experiment by Schmidt and collaborators~\cite{Schmidt00}, which induced fission reactions in various neutron-deficient actinides and pre-actinides between the At and U elements, represented a major breakthrough in the field. Since then, the R$^3$B/SOFIA collaboration has made great efforts to overcome the limitations of conventional fission experiments, enabling complete isotopic measurements of both fission fragments~\cite{JL2015_1,Pellereau17} through spallation, fragmentation, and Coulex reactions. 
This approach allowed to extract correlations between fission observables sensitive to the dynamics of the fission process~\cite{JLRS14,JLRS15_2,JLRS16_2} and the nuclear structure at the scission point~\cite{Chatillon19,Chatillon20,Martin15,Martin21,Chatillon22}. Recently, the existence of a proton shell $Z \approx 36$ in the region of pre-actinides has been clearly established with fission reactions of exotic nuclei measured by the R$^3$B/SOFIA collaboration~\cite{Morfouace24}, as shown in Fig.~\ref{fig:1} (right) together with data obtained from fusion-fission reactions~\cite{Mahata2022}. In the pre-actinide region the light fragment position remains fixed at $Z \approx 36$, and the heavy fragment charge increases steadily until it merges with the fission mode S1 at actinium. Thereafter, the fission modes S1 and S2 start to dominate, forcing the light fragment to increase its charge with the fissioning system charge (Z$_{CN}$). Here it is particularly noteworthy that the role reversal between the light and heavy fragments occurs precisely at the boundary separating pre-actinides from actinides.

With this proposal we aim at measuring the fission yields for neutron-deficient isotopes of Th, Ac and Ra, highlighted by the pink ellipse in Fig.~\ref{fig:1} (left), to improve our  understanding of the transition between pre-actinides and actinides. In particular, we seek to investigate the shell effects related to the neutron shell closure at $N=50$, which could indicate the existence of a new fission mode. The results from this study will offer essential data for the refinement of theoretical fission models~\cite{Ichikawa19,Scamps19,JLRS2022} in the region of the nuclear chart where pre-actinide and actinide isotopes intersect.

\section{Goals of proposed experiment}
The present R$^3$B experimental setup~\cite{JLRS_EPJ_23,Grana2023} permits to identify simultaneously both fission fragments in terms of their atomic and mass numbers. This approach made it possible to extract correlations between fission observables sensitive to the dynamics of the fission process~\cite{JLRS14,JLRS15_2,JLRS16_2,Grana2023,JLRS_EPJ_23_coulex} and the nuclear structure at the scission point~\cite{Chatillon19,Chatillon20,Martin15,Martin21,Chatillon22,Gabriel2023}. Here we will use Coulex-induced fission reactions of neutron-deficient isotopes of Th, Ac and Ra to search for the influence of the spherical neutron shell $N=50$ in the fission fragment yields~\cite{Scamps19}, as well as its dependence on excitation energy.

\newpage

\subsection{Searching for new neutron shells}
\begin{wrapfigure}{r}{8.5cm}
\centering  
\includegraphics[width=1.\linewidth]{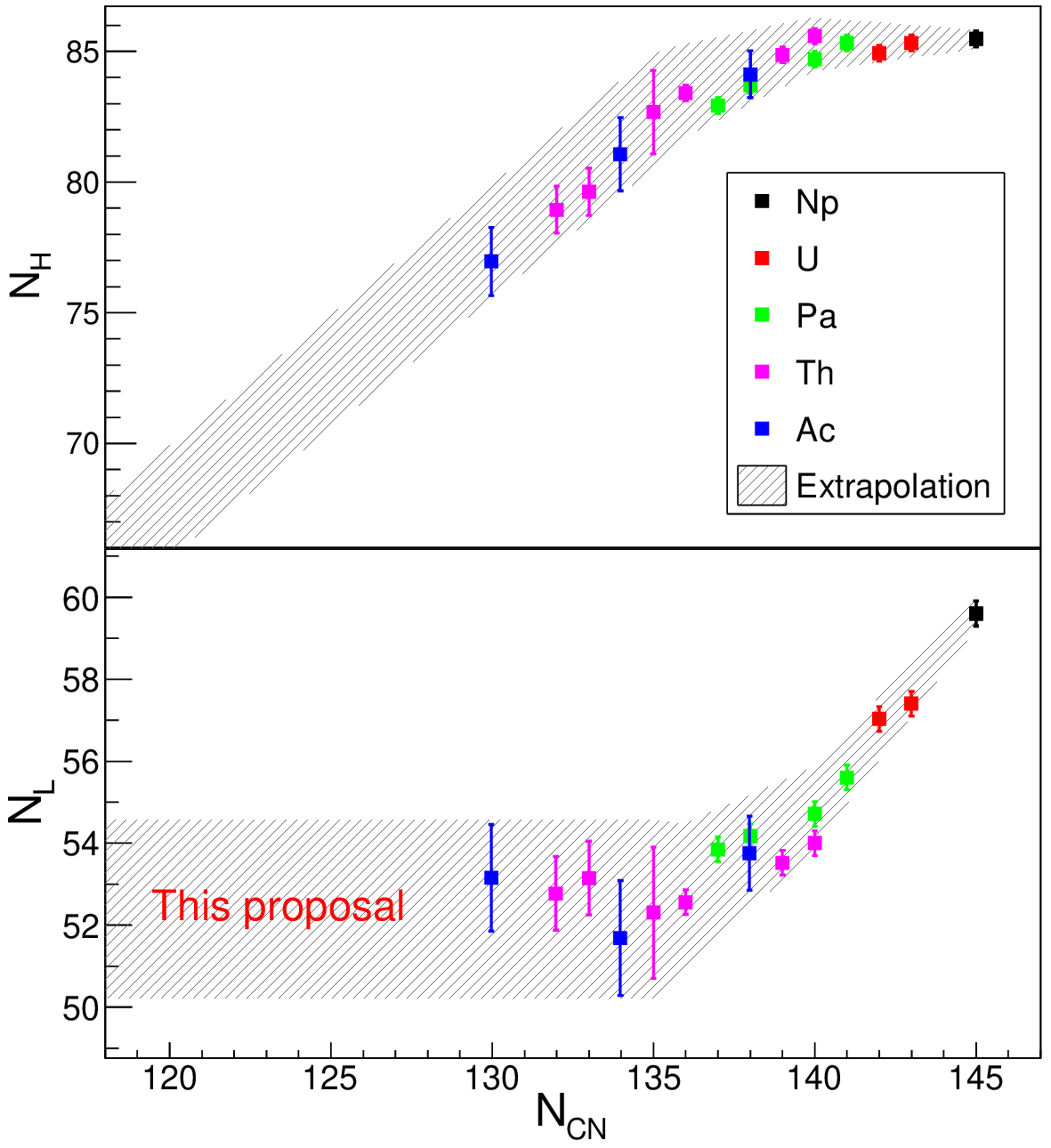}
\caption{Centroid positions of the light and heavy peaks of the isotonic yields measured prior the prompt-neutron evaporation phase~\cite{Chatillon22}. The shaded area represents an extrapolation of the data trend, assuming stabilization of the light fragment ($N_L$) between 50 and 54.}
\label{fig:2}
\end{wrapfigure}
The R$^3$B setup provides, on an event-by-event basis, an accurate determination of the fission yields in terms of charge and mass, as well as an accurate determination of the total prompt-neutron multiplicity for each pair of fission fragments~\cite{Martin21,Chatillon22}. From the first experiment performed in 2012~\cite{Pellereau17}, a total of 16 fissioning nuclei have been studied: six thorium isotopes $^{222,223,225,226,229,230}$Th spanning a wide range of $N/Z$ ratios, three actinium isotopes $^{219,223,227}$Ac, and seven heavier well-characterized actinides $^{228,229,231,232}$Pa, $^{234,235}$U, and $^{238}$Np. The fission fragment neutron yields of each fissioning nucleus can then be corrected for prompt-neutron evaporation to reconstruct the fission fragment neutron yields at the scission point. This allows for the study of shell effects in the potential-energy landscape of the fissioning system during the saddle-to-scission descent~\cite{Ichikawa12}. The fission yields can be fitted to determine the average positions of the various fission modes or their combined contributions. The results, shown in Fig.~\ref{fig:2}, present the average positions of the heavy and light fission fragment peaks as a function of the neutron number of the fissioning nucleus ($N_{CN}$). This systematic analysis confirms the well-established stabilization of the heavier fission fragment near the neutron shells $N=82-88$ for the heaviest fissioning nuclei $N_{CN}>139$. However, as the neutron number in the fissioning system decreases, the probability of populating the $N=82-88$ shells diminishes leading to a predominance of the $N=50-54$ neutron shells for the lighter fission fragment. This finding suggests a stabilization around $N=52$ that may indicate the presence of new fission modes. To confirm this, and the role of the spherical $N=50$ shell closure here, it is crucial to investigate the fission yields of lighter fissioning systems near $N_{CN}=126$ and determine whether this stabilization persists or vanishes.

\subsection{Fission yield dependence on excitation energy}
\begin{wrapfigure}{r}{8cm}
\centering
\includegraphics[width=0.9\linewidth]{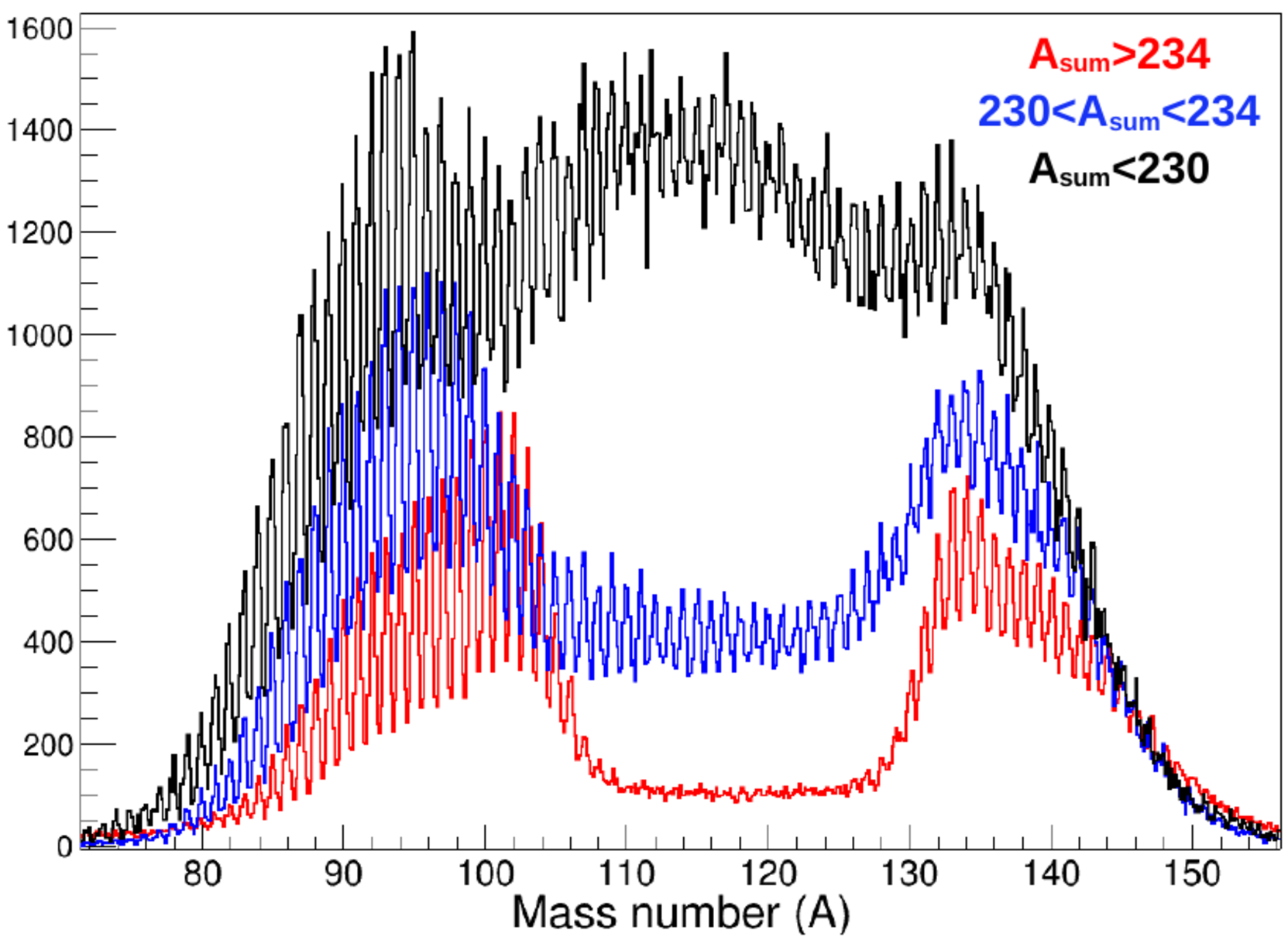}
\caption{Example of the evolution of the mass fission yields as a function of the sum of masses $A_1 + A_2$ for Coulex-induced fission of $^{238}$U~\cite{Pellereau17}.
}
\label{fig:3}
\end{wrapfigure}
The measurement of fission fragment masses provides a novel fission observable, the mass sum $A_1 + A_2$, which is sensitive to the initial excitation energy of the fissioning system~\cite{AntiaPhD}. 
This observable complements the total neutron emission, adhering to the conservation laws of neutrons and protons during the de-excitation process of the excited fissioning nucleus. This new observable can be used to suppress the symmetric fission mode in the fission yields by selecting values of $A_1 + A_2$ close to the mass of the initial fissioning nucleus~(compound nucleus). This allows us to select fission events with very small neutron evaporation and therefore much lower excitation energies. As shown in Fig.~\ref{fig:3}, for Coulex-induced fission reactions of $^{238}$U~\cite{Pellereau17}, applying the condition $A_{sum}=A_1 + A_2 > 234$ reduces the contribution of the symmetric fission channel, highlighting the asymmetric fission modes. This methodology allows for a more precise extraction of the positions of the light and heavy fission fragment peaks shown in the Fig.~\ref{fig:2}. It also enables the systematic study of the evolution of the fission yields with the excitation energy of the initial fissioning system and the neutron emission, in particular to understand the effects of neutron evaporation on the fission yields.

\section{Experimental approach}
\begin{figure}[h!]
    \centering
    \includegraphics[width=0.75\linewidth]{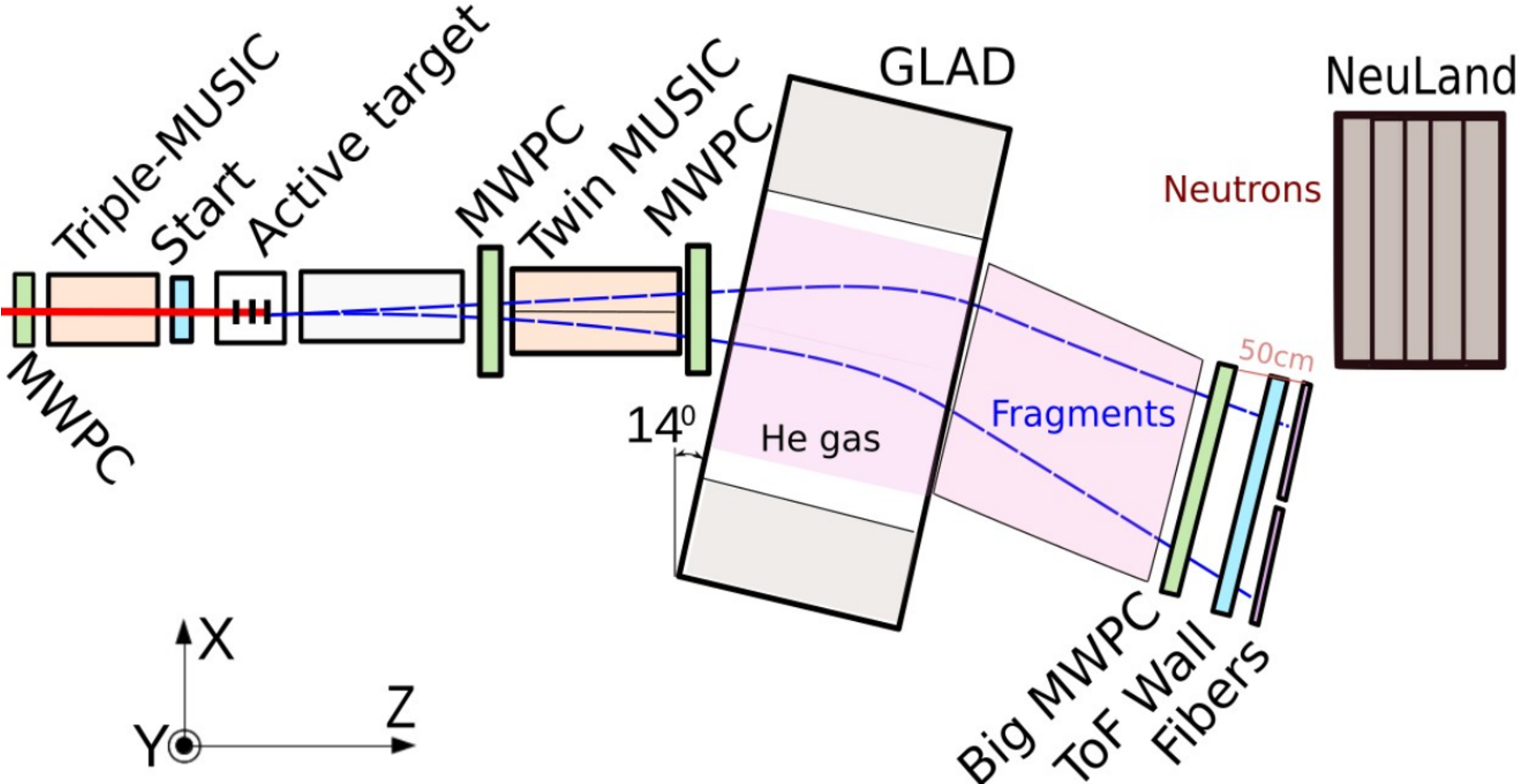}
    \caption{Schematic overview of the R$^3$B experimental setup at Cave-C.
    }
    \label{fig:4}
\end{figure}

The secondary beams will be produced with a primary beam of $^{238}$U at 1GeV/u impinging onto a Be target with a thickness of 1625 mg/cm$^2$ together with a 220.8 mg/cm$^2$ Nb stripper foil on the exit side. This setup will produce a high fraction of fully ionized secondary exotic projectiles that will be guided to the experimental area Cave-C to measure the fission reactions with the R$^3$B experimental setup, as shown in Fig.~\ref{fig:4}. It is divided into two parts, one to characterize the incoming projectile nuclei coming from the FRS and another to measure the fission products. The projectiles will be identified in terms of both charge and mass using the $\Delta$E-B$\rho$-ToF technique. To do so, a triple multiple-sampling ionization chamber~(triple-MUSIC) will be employed to determine the charge ($\Delta Z/Z = 2.5\times 10^{-3}$). Then the B$\rho$ will be derived from the positions measured at the FRS central focal plane and at the entrance of Cave-C with a multi-wire proportional counter~(MWPC), which combined with the time-of-flight~(ToF), provides the identification in mass ($\Delta A/A = 7\times 10^{-4}$)~\cite{Chatillon19,Martin21,Chatillon22}.

The second part consists of an active target, three MWPC detectors, a double multi-sampling ionization chamber (Twin MUSIC), a large acceptance superconducting dipole magnet (GLAD), a large ToF wall, and two fiber detectors. The Twin MUSIC chamber has a central vertical cathode that divides its volume into two active regions (left and right) both of which are divided into two sections (up and down). Each section is then segmented in 16 anodes that provide 16 independent energy-loss and drift-time measurements. This segmentation allows us to obtain the atomic number of our fission fragments with a resolution better than 0.34 charge units (FWHM) and the angles on the $X$-$Z$ plane with a resolution around 1 mrad (FWHM). MWPCs, situated in front and behind GLAD, provide the horizontal (\textit{X}) and vertical (\textit{Y}) positions of the fission fragments. The two MWPCs situated in front of the dipole magnet provides the \textit{X} and \textit{Y} positions with a resolution of around 200 $\mu$m and 1.5 mm (FWHM), respectively, while the MWPC situated behind GLAD provides these positions with a resolution of around 300 $\mu$m and 2 mm (FWHM), respectively. The ToF wall consists of 28 plastic scintillators that allow to measure the ToF of the fission fragments with respect to the start signal with a resolution of around 40~ps~(FWHM)~\cite{Ebran}. Finally, two fiber detectors will be located after the ToF-Wall to measure the $X$ position of the fission fragments with a resolution around 250~$\mu$m~(FWHM), which will allow us to determine the angles of the fission fragments.
This setup is expected to result in mass resolutions better than 0.8 mass units (FWHM)~\cite{JL2015_1,Pellereau17,Chatillon19,Chatillon20,Martin21}. Finally, the NeuLAND detector~\cite{Nueland}, equipped with 16 double scintillator planes, will be used to measure neutron multiplicities at low excitation energies.

The GLAD magnet will operate with a magnetic field of approximately 2~T, with its gap filled with helium gas at atmospheric pressure.
\begin{wraptable}{r}{6cm} 
    \centering
    \begin{tabular}{|c|c|}
        \hline
                         Isotope  & Fission events   \\ \hline
        \rowcolor{softyellow}  $^{213}_{123}$Th  & 830 \\ \hline
        \rowcolor{softyellow}  $^{214}_{124}$Th  & 1808 \\ \hline
        \rowcolor{softyellow}  $^{215}_{125}$Th  & 6029 \\ \hline
        \rowcolor{lightyellow}  $^{216}_{126}$Th  & 19292 \\ \hline
        \rowcolor{softyellow}  $^{209}_{120}$Ac  & 980 \\ \hline
        \rowcolor{softyellow}  $^{210}_{121}$Ac  & 3919 \\ \hline
        \rowcolor{softyellow}  $^{211}_{122}$Ac  & 7355 \\ \hline
        \rowcolor{lightyellow}  $^{212}_{123}$Ac  & 17888 \\ \hline
        \rowcolor{lightyellow}  $^{213}_{124}$Ac  & 49972 \\ \hline
        \rowcolor{lightyellow}  $^{214}_{125}$Ac  & 143561 \\ \hline
        \rowcolor{softyellow}  $^{207}_{119}$Ra  & 980 \\ \hline
        \rowcolor{softyellow}  $^{208}_{120}$Ra  & 19594 \\ \hline
        \rowcolor{lightyellow}  $^{209}_{121}$Ra  & 78374 \\ \hline
        \rowcolor{lightyellow}  $^{210}_{122}$Ra  & 211003 \\ \hline
        \rowcolor{lightyellow}  $^{211}_{123}$Ra  & 452140 \\ \hline
        \rowcolor{lightyellow}  $^{212}_{124}$Ra  & 789713 \\ \hline
        \rowcolor{lightyellow}  $^{213}_{125}$Ra  & 1066100 \\ \hline
    \end{tabular}
    \caption{Expected number of fission events for each isotope, with those shaded in darker color corresponding to isotopes being studied for the first time in this experiment. This estimation is for 6 main shifts of data taking with secondary beams.}
    \label{tab1}
\end{wraptable}
This setup is designed to minimize both the energy loss and angular straggling of the fission fragments as in previous experiments~\cite{JL2015_1,Pellereau17,Chatillon19,Martin21}. Additionally, a He-filled transfer pipe will be located between GLAD and the big MWPC to further mitigate these effects. 

The measurements will involve five FRS settings, each centered on the isotopes $^{212,213,214,215,216}$Th. These settings will allow the investigation of fission reactions for the isotopes listed in Table~\ref{tab1}. The total number of fission events expected to be measured with the R$^3$B experimental setup was calculated assuming an average $^{238}$U rate of $1.43\times10^8$ ions per second on the FRS target with a spill duration of 10~s (2~s extraction). To obtain the rate of secondary beams we have used the experimental production cross sections for the Th, Ac, and Ra isotopes coming from the fragmentation reaction $^{238}$U+Be at 1 GeV/u, as reported in Refs.~\cite{Taieb03, Junghans98}, and considering a Be target thickness of 1625~mg/cm$^2$. For the FRS transmission, an average factor of $10\%$ was assumed, which also represents the worst-case scenario. Finally, an average reaction rate of $9\%$ was considered for the active target made of Pb and C foils, with an average fission probability of $15\%$.

For the isotopes highlighted in darker color in Table~\ref{tab1}, which will be studied for the first time, we will measure the charge distribution since the statistics is enough for this purpose~\cite{Chatillon22}. For the remaining isotopes, the charge and mass distributions, together with TKEs, will be measured simultaneously. The fission yields obtained for these isotopes will serve to investigate the presence of the neutron shell $N=50$ and to study other fission observables, such as the even-odd effects~\cite{Steinh98,Jurado13,Moller2024} and the deformations at the scission configuration~\cite{Chatillon20,Bockstiegel08,Caa17}.

\section{Beam time request}
A $^{238}$U primary beam with an average intensity of $1.43\times10^8$~ions/s on the FRS target will be used to produce secondary beams of $^{213-217}$Th, $^{209-214}$Ac and $^{207-213}$Ra. The beam will be operated with a spill duration of 10s with a 2s extraction cycle. All required detectors for this experiment are fully operational and ready to perform the measurements. However, due to the relocation of the GLAD magnet from Cave-C to HEC, this experiment should be scheduled in 2026. We will need \textbf{6 main shifts for data taking with secondary beams} and \textbf{3 main shifts for data taking with $^{238}$U} that will be used for ion-tracking calibrations needed to reconstruct the fission fragment masses.

In \textbf{total}, we request \textbf{9 main shifts} of \textbf{$^{238}$U at 1~GeV/u} for data taking and \textbf{10 parasitic shifts} with the same beam for detector commissioning and associated calibrations. The parasitic beam time and 3 main shifts will be shared with the proposal titled "\textit{Exploring the new island of asymmetric fission in very neutron-deficient light fissioning system using the SOFIA@R$^3$B setup}". Therefore, if both proposals are approved, this proposal will require only \textbf{6 main shifts for data taking} with secondary beams.

\section{Acknowledgements}

The authors would like to thank H.~Alvarez-Pol, M.~Caama\~no-Fresco, A.~Chatillon, A.~Graña-González, A.~Heinz, E.~Haettner, and S.~Pietri for their valuable comments.

\begin{small}
\begin{spacing}{0.01} 

\end{spacing}
\end{small}


\begin{thebibliography}{20}

\bibitem{Schmidt2018}
K.-H.~Schmidt and B.~Jurado, Rep. Prog. Phys. \textbf{81}, 106301 (2018)

\bibitem{Aumann2024}
\underline{T.~Aumann et al., Phil.~Trans.~R.~Soc.~A~\textbf{382}, 20230121 (2024)}

\bibitem{Ichikawa19}
T.~Ichikawa and P.~M{\"o}ller, Phys. Lett. B~\textbf{789}, 679 (2019) 

\bibitem{Moller2015}
P.~M{\"o}ller and J.~Randrup, Phys. Rev. C \textbf{91}, 044316 (2015)

\bibitem{Scamps19}
G.~Scamps and C.~Simenel, Phys. Rev. C~\textbf{100}, 041602 (2019)

\bibitem{Schmidt00}
K.-H.~Schmidt et al., Nucl. Phys. A \textbf{665}, 221 (2000)

\bibitem{JL2015_1}
\underline{\textbf{J.~L.~Rodr\'{i}guez-S\'{a}nchez et al., Phys. Rev. C \textbf{91}, 064616 (2015)}}

\bibitem{Pellereau17}
\underline{\textbf{E.~Pellereau et al., Phys. Rev. C \textbf{95}, 054603 (2017)}}

\bibitem{JLRS14}
\textbf{J.~L.~Rodr\'{i}guez-S\'{a}nchez et al., Phys. Rev. C \textbf{90}, 064606 (2014)}

\bibitem{JLRS15_2}
\textbf{J.~L.~Rodr\'{i}guez-S\'{a}nchez et al., Phys. Rev. C \textbf{92}, 044612 (2015); Phys. Rev. C \textbf{94}, 034605 (2016)}


\bibitem{JLRS16_2}
\underline{\textbf{J.~L.~Rodr\'{i}guez-S\'{a}nchez et al., Phys. Rev. C \textbf{94} 061601(R) (2016)}}

\bibitem{Chatillon19}
\underline{\textbf{A.~Chatillon et al., Phys. Rev. C \textbf{99}, 054628 (2019)}}

\bibitem{Chatillon20}
\underline{\textbf{A.~Chatillon et al., Phys. Rev. Lett. \textbf{102}, 202502 (2020)}}

\bibitem{Martin15}
\underline{\textbf{J.-F.~Martin et al., Eur. Phys. J. A \textbf{51}, 174 (2015)}}

\bibitem{Martin21}
\underline{\textbf{J.-F.~Martin et al., Phys. Rev. C \textbf{104}, 044602 (2021)}}

\bibitem{Chatillon22}
\underline{\textbf{A.~Chatillon et al., Phys. Rev. C \textbf{106}, 024618 (2022)}}

\bibitem{Morfouace24}
\underline{\textbf{P.~Morfouace et al., Nature \textbf{641}, 339 (2025)}}

\bibitem{Mahata2022}
K.~Mahata et al., Phys. Lett. B \textbf{825}, 136859 (2022)

\bibitem{JLRS2022}
J.~L.~Rodr\'{i}guez-S\'{a}nchez et al., Phys. Rev. C \textbf{105}, 014623 (2022)

\bibitem{JLRS_EPJ_23}
\underline{\textbf{J.~L.~Rodr\'{i}guez-S\'{a}nchez et al., EPJ Web of Conf.~\textbf{284}, 04020 (2023)}}

\bibitem{Grana2023}
\textbf{A.~Graña-Gonz\'{a}lez et al., EPJ Web of Conf.~\textbf{290}, 02015 (2023)}

\bibitem{JLRS_EPJ_23_coulex}
\underline{\textbf{J.~L.~Rodr\'{i}guez-S\'{a}nchez et al., EPJ Web of Conf.~\textbf{290}, 02016 (2023)}}

\bibitem{Gabriel2023}
\textbf{G.~Garc\'{i}a-Jim\'{e}nez et al., EPJ Web of Conf.~\textbf{290}, 02009 (2023)}

\bibitem{Ichikawa12}
T.~Ichikawa et al., Phys. Rev. C~\textbf{86}, 024610 (2012)

\bibitem{AntiaPhD}
\textbf{A.~Graña-Gonz\'{a}lez, PhD dissertation, https://hdl.handle.net/10347/37672}

\bibitem{Ebran} 
A.~Ebran et al., Nucl. Instr. Meths. A \textbf{728}, 40 (2013)

\bibitem{Nueland}
K.~Boretzky et al., Nucl.~Inst.~Meth.~Phys.~Res.~A \textbf{1014}, 165701 (2021)

\bibitem{Taieb03}
J.~Taïeb et al., Nucl.~Phys.~A \textbf{724}, 413 (2003)

\bibitem{Junghans98}
A.~R.~Junghans et al., Nucl.~Phys.~A \textbf{629}, 635 (1998)

\bibitem{Steinh98}
S.~Steinh{\"a}user et al., Nucl.~Phys.~A \textbf{634}, 89 (1998)

\bibitem{Jurado13}
K.-H.~Schmidt and B.~Jurado, Phys. Procedia~\textbf{47}, 88 (2013)

\bibitem{Moller2024}
P.~M{\"o}ller and C.~Schmitt, Eur. Phys. J. A \textbf{60}, 27 (2024)

\bibitem{Bockstiegel08}
C.~B{\"o}ckstiegel et al., Nucl.~Phys.~A \textbf{802}, 12 (2008)

\bibitem{Caa17}
M.~Caamaño and F.~Farget, Phys. Lett. B \textbf{770}, 72 (2017)

\end{thebibliography}
\end{document}